\documentclass{cjaa}                   

\usepackage{graphicx}                   
\input{epsf.sty}                        
\input{psfig.sty}                       

\begin{document}

\title{Multiwavelength Observations of Ultraluminous X-Ray Sources}

\volnopage{Vol.0 (200x) No.0, 000--000}      
\setcounter{page}{1}          

\author{Philip Kaaret
\inst{}\mailto{philip-kaaret@uiowa.edu}}
\offprints{P. Kaaret}                   

\institute{Department of Physics and Astronomy, University of Iowa, 
           Van Allen Hall, Iowa City, IA, 52242, United States\\
           \email{philip-kaaret@uiowa.edu}
          }

\date{Received~~2004 September 27; Accepted }

\abstract{Multiwavelength observations may help us understand the
physical nature of the ultraluminous X-ray sources (ULXs) found in
external galaxies. Enabled by the arcsecond X-ray source positions now
available from Chandra, there has been significant recent progress in
the identification of optical and radio counterparts to ULXs.  Recent
results are reviewed on the identification of optical stellar
counterparts to ULXs, the relation between ULXs and regions of active
star formation, the study of optical nebular counterparts to ULXs,  and
the estimation of the luminosity of ULXs via their influence on
surrounding nebula.}


\authorrunning{P.\ Kaaret}            
\titlerunning{Multiwavelength Observations of ULXs}  

\maketitle


%
%

\section{Introduction}           

Ultraluminous X-ray sources (ULXs) are bright, variable X-ray sources
found in external galaxies and displaced from the galactic nucleus. 
Such sources were first discovered with the Einstein satellite
\cite{fabbiano89} and have been studied extensively with all subsequent
X-ray observatories.  The definition of ULX has been rather vague.  In
the interest of concreteness, I propose to establish the definition of
ULXs as irregularly variable, non-nuclear, X-ray sources with apparent
bolometric luminosities (estimated from the observed flux assuming
isotropic emission) exceeding the Eddington limit for a $20 M_{\odot}$
compact object.  Including only irregularly variable sources should
exclude most, if not all, young supernovae and supernova remnants,
leaving the accreting sources.  Because the masses of all of the black
hole candidates within our own Galaxy are below $20 M_{\odot}$
\cite{mcclintock03}, this definition would establish ULXs as
(observationally) distinct objects from the black hole X-ray binaries
found in the Milky Way.  Placing the luminosity cutoff at a lower value
would include objects similar to the black hole binaries in the Milky
Way.

The physical nature of the ULXs is very poorly constrained. 
Fundamental questions remain, even regarding the mass of the compact
objects.  Some of the major questions concerning ULXs are:

\begin{itemize}
  \item In what environments do ULX occur?
  \item Are there any ULXs in elliptical galaxies?
  \item Are there multiple classes of ULXs?
  \item Are the sources isotropic or beamed emitters?
  \item Do ULXs produce radio emission?
  \item What are the mass donors of ULXs?
  \item What are the properties of ULX binaries?
  \item How are ULXs formed?
\end{itemize}

In this review, we will examine these questions and how observations at
multiwavelengths can help understand the physical nature of the
ultraluminous X-ray sources.

\section{Models}

The great interest in ULXs arises from the possibility that they
represent a new class of black with masses intermediate between those
of stellar-mass black holes and supermassive black holes.  The main
motivation for this is the high inferred luminosities -- under the
assumption of isotropic emission -- (\cite{colbert99}).  Although the
spectral properties of the sources measured with the ASCA satellite
were also interpreted as additional evidence \cite{makishima00}. The
existence of intermediate mass black holes of significance because such
objects may require different formation mechanism than stellar-mass
black holes.  Madau \& Rees (2002) have suggested that such objects
could be the relicts of the first generation of stars, which are
thought to have had very low metallicity and therefore been extremely
massive. Intermediate mass black holes could also be formed by
interactions in stellar clusters (\cite{ebisuzaki01};
\cite{portegies04}).  The high stellar density within superstar
clusters found in starburst galaxies are a particularly appealing
location for the formation of intermediate mass black holes via
multiple stellar collisions.

However, alternative interpretations of the ULXs exist.  The difficulty
of forming intermediate mass black holes motivated King et al.\ (2001)
to suggest that the ULXs are, instead, mechanically beamed sources. In
their specific model, the mechanical beaming is produced by
super-Eddington accretion rates caused by thermal time scale mass
transfer of mass off a companion star.  The thermal time scale mass
transfer phase is suggest to be a common, but short lived, evolutionary
phase for young, high mass stars.

Another interpretation of the ULXs is that they may be relativistically
beamed jet sources, ``micro-blazars'', suggested to exist in external
galaxies after the discovery of microquasars in our own Galaxy
(\cite{mirabel99}).  Relativistic jets from microquasars have been
observed to produce X-ray as well as radio emission (\cite{corbel02}). 
K\"ording et al.\ (2001) suggest that beaming can produce the intense
X-ray fluxes observed from ULXs.  One possible mechanism which have
been studied in some detail is inverse-Compton interactions of photons
from a high mass companion star with the energetic particles in a
relativistic jet (\cite{geo02}).

Another possibility is that the sources are not beamed, but
super-Eddington radiators.  Begelman (2001) has suggested the potential
existence of accretion flows in which the Eddington limit is
significantly violated.  The basic mechanism is that radiation-pressure
dominated disks are highly inhomogeneous due to photon bubbles, and,
therefore, may be able to radiate at rates higher than predicted in
standard accretion disk theory -- rates up to 10--100 times the
Eddington luminosity.

\section{Multiwavelegth observations}

Each of these different models leads to different predictions for the
beaming factor of the X-ray emission and therefore the underlying
number of sources required to produce the observed ULX population, for
the mass of the black hole and the evolutionary state of the companion
star and therefore the formation mechanism of the black hole and binary
and the nature of the stellar populations where ULXs should be found,
and for the duty cycle of the X-ray emission.

The motivation for performing multiwavelength observations comes from
the history of X-ray astronomy and the realization that, with the
possible exception of X-ray pulsars, the identification and study of
counterparts to X-ray sources at other wavelengths has been necessary
to understand the physical nature of all known classes of X-ray
sources.  Multiwavelength observations should help distinguish between
the models described in the previous section.  Specifically,

\begin{itemize} 

  \item The study of the stellar environments of ULXs will allow us to
  determine what types of stars are their progenitors (i.e.\ young or
  old) and determine if any specific conditions are required for the
  occurrence of ULXs (low metallicity, high star formation density). 
  This knowledge will help constrain models of the formation mechanisms
  of ULXs.

  \item Identification of stellar companions and determination of the
  spectral types of the companions will directly constrain the
  evolutionary history of the binaries containing ULXs.  Some of the
  models described above make very specific predictions concerning the
  type of the companion star which can be tested.

  \item Spectroscopy of companion stars could enable measurement of
  radial velocity curves which would provide direct constraints on the
  compact object mass.  This possibility is very exciting and could
  lead to definitive evidence for or against the interpretation of ULXs
  as intermediate-mass black holes.

  \item Determining whether some, all, or none of the ULXs produce
  radio emission should help determine whether or not relativistic
  beamed jet models for the ULXs are viable.  Radio emission is an
  essential feature of relativistic jets and should be present if the
  ULXs are, indeed, micro-blazars.

  \item The identification and study of nebula associated with ULXs may
  enable us to probe the origin of the ULXs, if the nebula are
  associated with the birth event in a manner similar to supernova
  remnants, or enable us to determine the total radiation and particle
  fluxes from the ULXs.  Study of nebula will help constrain the
  formation mechanisms and the total energetics of ULXs.

\end{itemize}

The multiwavelength study of ULXs is a field which is just now blooming
because of the accurate X-ray positions provided by the Chandra X-ray
Observatory.  The precise astrometry possible with Chandra has enable,
for the first time, the unique identification of optical counterparts
to ULXs.  In the next several sections, we present a review of some
recent results in multiwavelength observations of ULXs.

\section{Optical counterparts in star-forming galaxies}

ULXs are found preferentially in actively star-forming galaxies and the
ULXs in star-forming galaxies include the brightest and most highly
variable members of the class.  The ULXs in star-forming galaxies tend
to be spatially coincident with the regions of active star formation, a
trend which is discussed qualitatively below.  This trend also makes
the identification of unique optical counterparts difficult because ULX
fields are crowded when imaged in the optical.

Pre-Chandra, progress in identification of optical counterparts was
minimal because of the large number of stars in each Rosat error
circle.  With Chandra, the situation has improved greatly, but remains
difficult.  An illustrative example is given by the ULX in NGC 5204.
Goad et al.\ (2002) analyzed an HST image of the field of the ULX and
considered objects found in the relative HST/Chandra error circle. The
size of the error circle was dominated by the absolute astrometric
accuracy of Chandra.  Unfortunately, they found three potential
counterparts. Measuring the magnitudes and colors of the counterparts,
they found a range of objects including a single F2-F5 supergiant, 2-3
A2 supergiants, 2-3 B5 supergiants, or a small and young stellar
cluster.  There are more possibilities than counterparts because the
magnitudes and colors of some of the counterparts could be interpreted
in multiple ways.

The situation can be significantly improved if objects, other than the
ULX, can be identified in both HST and Chandra images.  In this case,
the uncertainty in relative astrometry can be greatly decreased as
compared with the uncertainty in the absolute astrometry of Chandra or
HST.  However, the number of cases where this has been possible as
small.  The major hurdle at this point is the low number of X-ray
sources in the subset of the Chandra fields overlapping the HST field.
Liu et al.\ (2002) used SN 1993J to align HST and Chandra images of M81
and obtain relative astrometry accurate to $0.2\arcsec$ and identify a
unique optical counterpart to the ULX NGC 3031 X-11.  Zampieri et al.\
2003 used SNR in NGC 1313 for relative astrometry to identify
counterpart to NGC1313 X-2 (figure above).  Liu et al.\ (2004) used the
coincidence of an optical bright object with a Chandra source to
improve on the astrometry for NGC 5204 and unique identify the ULX with
one of three possible counterparts found by Goad et al.\ (2002). The
fourth unique optical counter part identification was made for the ULX
i Holmberg II where the presence of a bright He{\sc ii} nebula and a
relatively uncrowded field made a unique counterpart identification
possible without highly accurate relative astrometry.

\begin{table}[tb]
\begin{center}
\begin{tabular}{lcc}
\hline
Source        & Counterpart       & Reference  \\ \hline
M81 X-6       & O8V               & Liu et al.\ 2002  \\
NGC 1313 X-2  & early O V, O-B I  & Zampieri et al.\ 2004 \\
Holmberg II   & O4V to B3 Ib      & Kaaret et al.\ 2004 \\
NGC 5204      & B0 Ib             & Liu et al.\ 2004 \\
\end{tabular}
\end{center}
\caption{Optical stellar counterparts to ULXs.  The table includes the
ULX name, the spectral type of the optical counterpart, and the
publication were the optical identification was reported.}
\label{Table:optids}
\end{table}

The current full set of unique optical counterpart identifications for
ULXs is given in Table~\ref{Table:optids}.  There are only four. 
However, so strong trends emerge even from this limited sample.  The
colors and magnitudes of the counterparts are consistent with young,
massive stars, in particular O or B stars.  This suggests that the ULXs
are very young objects.  However, care must be taken in interpretation
of the optical spectral types because reprocessed disk emission may
contribute to optical light.  The optical colors of low-mass X-ray
binaries in the Milky Way where the optical light is dominated by
reprocessed emission from the accretion disk are similar to those found
for the ULXs.

Optical spectroscopy of the counterparts should enable detailed
understanding of their nature.  The first step has been taken by Liu et
al.\ (2004) who obtained an HST/STIS far-ultraviolet spectrum of the
counterpart of the NGC 5204 ULX.  The line spectrum bolsters their
identification of the star as B0 Ib.  The spectrum also shows a N{\sc
V} emission line.  This is a very high ionization line which is not
seen in B star spectra, but is seen in X-ray illuminated accretion
disks and coronae.  The line is very suggestive that the correct
companion to the ULX has been identified.  Liu et al.\ (2004) suggest
that the B0 star fills it Roche lobe and that mass transfer proceeds
via Roche lobe overflow.  This mode of mass transfer differs from the
wind-fed accretion typical of high-mass X-ray binaries in the Milky Way
and was first suggested by Kaaret et al.\ (2004) as necessary in order
to produce the high luminosities seen from ULXs.  The detection of
spectral lines from the optical companion to a ULX raises the exciting
possibility that it may be possible to obtain a radial velocity curve
and therefore dynamical constrains on the compact object mass.

\begin{figure*}
\centerline{\psfig{figure=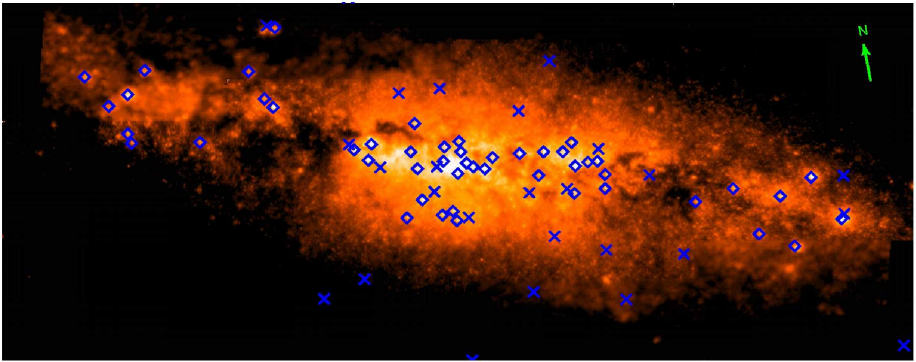,width=4.8in}}
\caption{Infrared image of M82 from HST/NICMOS with super star clusters
are marked as diamonds and X-ray sources are marked as X's (following
Kaaret et al.\ 2004).}
\label{Fig:m82ssc}
\end{figure*}

\section{ULX and super star clusters}

ULXs preferentially occur in starburst galaxies and the most luminous
ULXs are often found near sites of active star formation
\cite{zezas02}.  In starburst galaxies, a substantial fraction of young
stars are found in ``super star clusters'' -- luminous, compact,
clusters containing up to $10^{6} M_{\odot}$ of stars within a radius
of a few parsecs \cite{meurer95}.  Stellar encounters in such dense
clusters may lead to enhanced production of binaries, particularly
binaries containing compact objects \cite{portegies04}. 

\begin{figure}
\centerline{\psfig{figure=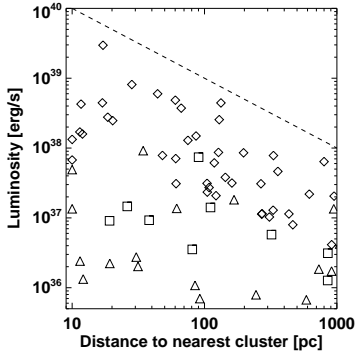,width=2.0in}}
\caption{Displacements of X-ray sources from the nearest super star
clusters for three nearby starburst galaxies: M82 (diamonds), NGC 1569
(triangles), NGC 5253 (squares)  (from Kaaret et al.\ 2004).  The
luminosities are calculated from the flux for the 0.3--8~keV band
assuming isotropic emission.  The dashed line represents the equation
$L_x = (1 \times 10^{41} \, {\rm erg \, s^{-1}}) / (d/{\rm pc})$.}
\label{Fig:dlum}
\end{figure}

Kaaret et al.\ (2004a) studied the spatial offsets between  super star
clusters and X-ray sources (both ULXs and normal X-ray sources) in
starburst galaxies, see Figures \ref{Fig:m82ssc} and \ref{Fig:dlum}. 
The position of the X-ray sources are well correlated with, but have
significant offsets from, the super star clusters.  Further, the
brighter X-ray sources preferentially occur closer to clusters. 
Because the star clusters are very good tracers of star formation
activity, the good correlation of X-ray sources with super star
clusters indicates that the X-ray sources are young objects associated
with current star formation.  This may suggest that the X-ray sources
and the ULXs in particular are produced via dynamical interactions in
the super star clusters \cite{portegies04}.

The offset of the X-ray sources from the super star clusters suggests
motion of the X-ray sources.  Such motions can naturally be produced by
dynamical interactions within the clusters leading to ejection of X-ray
binaries with speed comparable to the cluster escape velocity.  The
absence of very bright sources at large displacements from clusters may
help constrain models of the sources.  It particular, it suggests that
the ULXs are not simply a subset of the normal X-ray population viewed
at particular beaming angles.  Further, it suggests that the turn-on of
a ULX occurs promptly and is not significantly delayed after the
creation of the binary.

\section{Optical counterparts in non star-forming galaxies}

The presence of ULXs in early-type galaxies in which there is no active
star formation and the youngest stellar populations are billions of
years old, is an actively pursued question.  Irwin et al.\ (2003)
showed that the numbers of seeming ULXs in early-type galaxies are
consistent with expected number of background AGN.  This would suggest
there are no ULXs (meeting the definition above) in early-type
galaxies.  However, positive identifications of ULXs with globular
clusters identified in the optical can exclude a potential AGN
identification. The identification of cluster counterparts to ULXs in
NGC 1399 (Angelini et al.\ 2001) and NGC 4565 (Wu et al. 2002) show
that the total (isotropic equivalent) X-ray luminosities of some
globular clusters can be as high as $5\times 10^{39} \rm \, erg \,
s^{-1}$.  However, these luminosities may represent the summed output
from several different objects and no ULXs with (isotropic equivalent)
luminosities exceeding $10^{40} \rm \, erg \, s^{-1}$ have been found.

Intermediate mass black holes could be formed by dynamical interactions
in clusters, similar to the case described above for super star
clusters.  Such black hole could remain within the clusters and then
capture a star to become active again at the current epoch. 
Alternative explanations for the ULXs in globular clusters are that the
Eddington limit is violated, the emission is mechanically beamed, or
the emission is relativistically beamed (see section 2).  Kalogera et
al.\ (2004) suggest intermediate mass black holes will be transient,
while thermal time scale mass transfer sources will be persistent.
Transient behavior would then be a signature of an intermediate mass
black hole.  However, the time scale of the transient behavior is
expected to be quite long, ``far longer than an observer's lifetime''
\cite{kalogera04}.  In addition, relativistically beamed sources could
also be transient, which might make the interpretation of transient
behavior ambiguous.

\section{Nebular counterparts}

Optical and radio nebulae have been found which are spatially
coincidence and likely physically related to ULXs.  Such nebula may
have been produced in the birth event of the ULXs or may be continuous
powered by the ULX.  The study of nebulae associated with ULXs may
constrain epoch and energetics of the origin of ULX.  Study of the
nebulae also may constrain the current energetics of the ULXs and  help
determine if the X-rays observed directly are beamed or unbeamed, The
study of optical nebulae associated with ULXs has been pioneered by
Manfred Pakull and much of the state of our current knowledge is
summarized in Pakull \& Mirioni (2002).

While the number of optical nebulae associated with ULXs is relatively
small, the nebulae identified to date show some interesting
similarities.  Most of the nebula show emission line ratios similar to
those seen in supernova remnants.  This suggests that shocks powers the
nebula.  In analogy with standard supernova remnants, a natural
interpretation of these nebula is in terms of expanding shells of
material powered by an initial explosion.  In the case of NGC 1313 X-2,
high resolution spectroscopy of the emission lines provides direct
evidence for  expansion with a velocity of 80~km/s.

The nebula tend to be quite large, ranging in diameter from 200~pc for
the nebula surrounding IC 342 X-1 \cite{roberts03} to 400~pc for the
nebula near NGC 1313 X-2 (Pakull \& Mirioni 2002).  If the nebulae are,
indeed, expanding from a single initial explosion, then the large
diameters require very energetic explosive events with total energies
$\sim 10^{52} \rm erg$.  This is more energetic than a single
supernova.  Such energies can be produced via hypernova, potentially
linking the ULXs with gamma-ray bursts, or multiple supernovae.  The
dynamics of the nebula, in this case, would also require very young
ages for the ULXs, typically less than $1$~Myr.

An alternative interpretation of the nebulae is that they are
continuously energized by jets produced by the ULX.  The nebula would
then be similar to that surrounding the Galactic jet source SS 433. In
this case, the nebulae should reflect the total energy output of the
ULX.

\section{A black hole calorimeter}
 
Nebula powered by photoionization provide a potential means to measure
the total energy output of a ULX in all directions.  This would enable
us to answer the question or whether or not the X-rays from the ULX are
beamed along our line of sight.

\begin{table}[tb]
\begin{center}
\begin{tabular}{lccccc}
\hline
Observation & $L_{\rm X}$           & $L_{\rm PI}$         \\
            & [erg s$^{-1}$]        & [erg s$^{-1}$]       \\ \hline
10 April    & $16 \times 10^{39}$   & $5.9 \times 10^{39}$ \\
16 April    & $17 \times 10^{39}$   & $6.1 \times 10^{39}$ \\
19 Sept     & $ 5 \times 10^{39}$   & $3.7 \times 10^{39}$ \\ \hline
\end{tabular}
\end{center}
\caption{X-ray and photoionization luminosities for 3 observations of
Holmberg II.  The X-ray luminosity is the equivalent isotropic
luminosity calculated from the X-ray flux and spectrum measured using
XMM-Newton on the date indicated.  The photoionization luminosity is the
that required to produce the observed He{\sc ii} luminosity using the
X-ray spectrum measured with XMM-Newton on the data indicated.}
\label{Table:holmbergiilum}
\end{table}

The idea is that X-rays from the ULX will ionize the nebula.  Excited
atoms in the nebula will then produce line emission from high
excitation states.  A particularly useful emission line is the $\lambda
4686$ line from fully ionized Helium.  The He{\sc ii} emission is
proportional to the total luminosity of ionizing radiation from 54~eV,
the ionization threshold of He, to about 300~eV, and at most one He{\sc
ii} photon is produced for reach X-ray in this band.

Pakull \& Mirioni (2002) discovered a He{\sc ii} emission line nebula
near the ULX in the galaxy Holmberg II using ground-based optical
spectroscopy.   Using a thermal bremsstrahlung spectrum fit to
non-simultaneous Rosat and ASCA data and then folding the fitted
spectrum through a photoionization code to relate the He{\sc ii}
luminosity to the X-ray luminosity, they found that the total X-ray
luminosity required to produce the observed photoionization was in the
range $L_{\rm PI} = 3-13 \times 10^{39} \rm \, erg \,s^{-1}$.  The
dominant uncertainty in their luminosity estimate is the  He{\sc ii}
luminosity to X-ray luminosity conversion and is caused by the
uncertainty in the X-ray spectrum.

Kaaret, Ward, \& Zezas (2004) observed the same nebula using HST and
obtained images in the optical emission lines He{\sc ii} $\lambda
4686$, H$\beta$, and [O{\sc i}] $\lambda 6300$.  Examination of the
relative morphology of the nebula in the three lines shows that the
structure of nebula is consistent with that expected from
photoionization.  This strengthens the suggestion by Pakull \& Mirioni
(2002) that the nebula is photoionization powered.  The total He{\sc
ii} from the HST data is $2.7 \times 10^{36} \rm \, erg \, s^{-1}$.
Using archival XMM-Newton data, the X-ray spectrum of the ULX was fit
to a Comptonization model in which the seed photons for the
Comptonization are drawn from a multicolor disk black body spectrum.
The use of a Comptonization model rather than a simple powerlaw is
motivated because a model in which the low energy extension of the
spectrum is well defined, and preferentially physically motivated, is
needed for the photoionization modeling.  The fitted spectral model
from three different XMM-Newton observations were then used an inputs
to modeling of the photoionization nebula to calculate the conversion
factor between He{\sc ii} luminosity and X-ray luminosity.  The
estimated X-ray luminosity is at least $4 - 6 \times 10^{39} \rm \, erg
\, s^{-1}$, see Table~\ref{Table:holmbergiilum}.   The range in
luminosity comes from the different spectral fits for the different
XMM-Newton observations.  This is a lower bound on the true luminosity
because the HST images reveal that the nebula only partially covers the
ULX.  If the Eddington limit holds, then the implied minimum black hole
mass for the ULX is $25 M_{\odot}$.  This is greater than the measured
mass of any stellar black hole in our Galaxy and establishes the source
as truly ultraluminous (but not necessarily as an intermediate mass
black hole).

\section{Conclusions}
 
Multiwavelength study of ultraluminous X-ray sources is a very active
field and has led to significant new information about the nature of
the objects.  First, ULXs tend to be associated with young, high mass
stars and, in some cases, super star clusters.  This clearly indicates
that most ULXs are young objects and the association of ULXs with super
star clusters may suggest an origin via dynamical interactions in
clusters for at least part of the ULX population.  While high
luminosities have been observed from some globular clusters, the
incidence of ULXs in early-type galaxies has not been definitively
proven (strong variability from a cluster source would do so) and the
number of ULXs in early-type galaxies may be consistent with zero.

The study of optical nebulae associated with ULXs reveals unusually
large nebula, compared with normal supernova remnants, which show
evidence for shock emission.  The most natural interpretation of these
nebula is as remnants of explosions.  These explosions must have been
significantly more powerful than standard supernovae.  Alternatively,
the nebula may be continuous powered by jets emitted by the ULXs.
Finally, the energetics of the He{\sc ii} emission line nebula in
Holmberg II shows that at least one ULX is truly ultraluminous.

\begin{acknowledgements}

I thank Manfred Pakull, Martin Ward, Andreas Zezas, and Ed Colbert for
useful discussions. 

\end{acknowledgements}

\label{lastpage}

\end{document}